\documentclass[a4paper,preprintnumbers,amsmath,amssymb,nofootinbib,10pt]{revtex4}
\usepackage[margin=2cm]{geometry}
\usepackage{amssymb}
 \usepackage{epsfig}

\def\beq{\begin{equation}}
\def\eeq{\end{equation}}
\def\baq{\begin{eqnarray}}
\def\eaq{\end{eqnarray}}

\def\bea{\begin{eqnarray}}
\def\eea{\end{eqnarray}}
\def\be{\begin{equation}}
\def\ee{\end{equation}}

\def\Tvac{\check{T}}
\def\rhovac{\check{\rho}}
\def\Pvac{\check{P}}
\def\uvac{\check{u}}
\def\vvac{\check{v}}
\def\thetavac{\check{\theta}}

\def\Svac{\check{S}}

\def\de{{\rm de}}
\def\gcg{{\rm gCg}}

\begin{document}

%\title{\Large Inhomogeneous vacuum energy}
\title{Inhomogeneous and interacting vacuum energy}

%\bigskip

\author{Josue De-Santiago$^{\,a,b}$}
\author{David Wands$^{\,a}$}
\author{Yuting Wang$^{\,a,c}$}
 \hskip0.5cm\\
%\hskip1cm
%
\affiliation{$^a$Institute of Cosmology $\&$ Gravitation, University of Portsmouth, Dennis Sciama Building,
%CQG \\\hskip0.2cm
Portsmouth, PO1 3FX, United Kingdom}
% \\% \hskip0.2cm
\affiliation{$^b$Universidad Nacional Aut\'onoma de M\'exico, 04510, D.~F., M\'exico}
%Departamento de Fisica, Instituto Nacional de Investigaciones Nucleares, M\'{e}xico
%\\% \hskip0.2cm
\affiliation{$^c$School of Physics and Optoelectronic Technology, Dalian University of Technology, Dalian,
%CQG \\\hskip0.2cm
Liaoning 116024, People's Republic of China}
%\\
%\hskip0.1cm

%E-mail: \email{david.wands@port.ac.uk}}
\date{\today}

\begin{abstract}
Vacuum energy is a simple model for dark energy driving an accelerated expansion of the universe. If the vacuum energy is inhomogeneous in spacetime then it must be interacting. We present the general equations for a spacetime-dependent vacuum energy in cosmology, including inhomogeneous perturbations. We show how any dark energy cosmology can be described by an interacting vacuum+matter. Different models for the interaction can lead to different behaviour (e.g., sound speed for dark energy perturbations) and hence could be distinguished by cosmological observations. As an example we present the cosmic microwave microwave background anisotropies and the matter power spectrum for two different versions of a generalised Chaplygin gas cosmology.
% (Invited talk presented at the IVth International Meeting on Cosmology and Gravitation, Guadalajara, May 2012.)
\end{abstract}

\maketitle

%\abstract{..}
%\begin{abstract}...
%\end{abstract}

\section{Introduction}

Vacuum energy provides a very simple model of dark energy \cite{Copeland:2006wr,Bertolami:1986bg,Freese:1986dd,Chen:1990jw,Carvalho:1991ut,Berman:1991zz,Pavon:1991uc,AlRawaf:1995rs,Shapiro:2000dz,Buchert:2010ug,Sola:2011qr,Wands:2012vg}. It is the energy density that remains in the absence of any particles, and therefore it remains undiluted by the cosmological expansion. A positive vacuum energy can drive an accelerated expansion once the density of ordinary matter or radiation becomes sub-dominant. Unlike other models of dark energy it does not necessarily introduce any new dynamical degrees of freedom.

We define a vacuum energy, $V$, to have an energy-momentum tensor proportional to the metric
\be
 \label{Tvac}
 \Tvac^\mu_\nu = - V g^\mu_\nu  \,.
 \ee
By comparison with the energy-momentum tensor of a perfect fluid
\be
 \label{Tmunu}
 T^\mu_\nu = P g^\mu_\nu +  (\rho+P)u^\mu u_\nu \,,
 \ee
we identify the vacuum energy density and pressure with $\rhovac=-\Pvac=V$, but since there is no particle flow then the four-velocity of the vacuum, $\uvac^\mu$, is undefined.

A vacuum energy that is homogeneous throughout spacetime, $\nabla_\mu V=0$, is equivalent to a cosmological constant in Einstein gravity, $\Lambda=8\pi G_N V$. The discrepancy between the value of the energy density required by current observations and the typical energy scales predicted by particle physics is the long-standing cosmological constant problem \cite{Weinberg:1988cp}.

We will consider the possibility of a time and/or space dependent vacuum energy. {}From Eq.~(\ref{Tvac}) we have
 \be
\label{vac-conservation}
 \nabla_\mu \Tvac^\mu_\nu = Q_\nu \,.
 \ee
where the energy flow is given by
\be
 Q_\nu \equiv - \nabla_\nu V \,.
 \ee
We can therefore identify an {\em inhomogeneous vacuum}, $\nabla_\mu V\neq0$, with an {\em interacting vacuum}, $Q_\nu\neq 0$ \cite{Wands:2012vg}. The conservation of the total energy-momentum (including matter fields and the vacuum energy) in general relativity
\be
 \nabla_\mu \left( T^\mu_\nu + \Tvac^\mu_\nu \right) = 0 \,,
 \ee
implies that the vacuum transfers energy-momentum to or from the matter fields
\be
 \nabla_\mu T^\mu_\nu = -Q_\nu
 % = \nabla_\nu V
  \,.
 \ee

Note that the energy density and four-velocity of a fluid can be identified with the eigenvalue and eigenvector of the energy-momentum tensor (\ref{Tmunu}):
\be
 T^\mu_\nu u^\nu= -\rho u^\mu \,.
 \ee
Because the vacuum energy-momentum tensor (\ref{Tvac}) is proportional to the metric tensor, any four-velocity, $u^\mu$, is an eigenvector
\be
 \Tvac^\mu_\nu u^\nu= - V u^\mu \quad \forall \ u^\mu \,,
 \ee
and all observers see the same vacuum energy density, $V$, i.e., the vacuum energy is boost invariant.

Although the vacuum does not have a unique four-velocity, we can use the energy flow, $Q_\nu$, to define a preferred unit four-vector in an inhomogeneous vacuum  \cite{Wands:2012vg}
\be
 \label{def-uvac}
 \uvac^\mu = \frac{-\nabla^\mu V}{|\nabla_\nu V\nabla^\nu V|^{1/2}} \,.
 \ee
normalised such that $\uvac_\mu\uvac^\mu=\pm1$ for a spacelike or timelike flow. Note however that the $\uvac^\mu$ defines a potential flow, i.e, with vanishing vorticity.

In this paper we will consider vacuum energy which may be inhomogeneous in spacetime, interacting with fields or fluids without necessarily invoking additional degrees of freedom. We show that any spatially homogeneous dark energy cosmology can be decomposed into an interacting vacuum+matter cosmology. By lifting this spatially-homogeneous solution to a covariant interaction one can study inhomogeneous perturbations which obey coupled first-order equations of motions for the matter density and velocity.
% We present the equations of motion for linear perturbations in Einstein gravity and identify gauge-invariant variables for the vacuum perturbations.
As an example we consider perturbations of a generalised Chaplygin gas cosmology, decomposed into an interacting vacuum+matter.

\section{Vacuum cosmology}

\subsection{FRW background}

The symmetries of a spatially homogeneous and isotropic Friedmann-Robertson-Walker (FRW) metric, with scale factor $a(t)$ and Hubble rate $H=\dot{a}/a$, require the vacuum to be spatially homogeneous and isotropic too, hence $V=V(t)$. In this case the vacuum and matter are both homogeneous on spatial hypersurfaces orthogonal to the matter four-velocity, $u^\mu=(1,0,0,0)$. Note that the energy flow, $\uvac^\mu$, and matter velocity, $u^\mu$, necessarily coincide in FRW cosmology due to the assumption of isotropy.

The Friedmann constraint equation requires
\be
 H^2 = \frac{8\pi G_N}{3} \left( \rho + V \right) - \frac{K}{a^2} \,,
 \ee
where $K$ determines the spatial curvature.
The continuity equations for matter fields and vacuum are
\bea
 \dot\rho + 3H(\rho+P) &=& -Q
\,,\\
 \dot{V} &=& Q
\,.
\eea
The vacuum energy, $V$, is undiluted by the cosmological expansion, but can have a time-dependent density in the presence of a non-zero energy transfer, $Q\neq 0$, where
\be
 Q \equiv -u^\nu Q_\nu \,.
 \ee

There have been many attempts to describe the present acceleration as due to a time-dependent vacuum energy
% v3
\cite{Bertolami:1986bg,Freese:1986dd,Chen:1990jw,Carvalho:1991ut,Berman:1991zz,Pavon:1991uc,AlRawaf:1995rs,Shapiro:2000dz,Buchert:2010ug,Sola:2011qr}. %
However, there is little to be learnt from simply assuming an arbitrary time-dependent vacuum to obtain the desired cosmological solution. Ideally one should have a physical model from which one can derive a time-dependent solution and study other physical effects.
% , including the evolution of inhomogeneous perturbations which can be tested against other cosmological observations.
%
For example, vacuum fluctuations of free fields can support an averaged density proportional to the fourth-power of the Hubble expansion, $V\propto H^4$ \cite{Bunch:1978yq}. Such a vacuum energy would not in itself support an accelerated expansion, but other forms such as $V\propto H$ have been proposed \cite{Schutzhold:2002pr} better able to match the observational data \cite{Fabris:2006gt,VelasquezToribio:2009qp,Zimdahl:2011ae,Xu:2011qv,Alcaniz:2012mh}.

It is not obvious how such time-dependent vacuum models can be compared with observations in an inhomogeneous universe. We will argue that it is possible to give a consistent description of vacuum dynamics, and in particular the relativistic equations of motion for inhomogeneous perturbations, given a covariant, physical prescription for the local vacuum energy or, equivalently, the vacuum energy transfer 4-vector, $Q_\mu=-\nabla_\mu V$. One should then be able to subject vacuum models to observational constraints, even in the absence of a Lagrangian derivation or microphysical description.

% \subsection{Covariant 1+3 split}

\subsection{Linear perturbations}

Let us consider inhomogeneous linear, scalar perturbations where the energy and pressure of matter is given by $\rho(t)+\delta\rho(t,x^i)$ and $P(t)+\delta P(t,x^i)$, and the four-velocity of matter is given by
\be
 \label{u}
 u^\mu = \left[ 1-\phi \,, a^{-1}\partial^i v \right] \,, \quad u_\mu = \left[ -1-\phi \,, \partial_i \theta \right] \,.
 \ee
where we define $\partial^iv=a(\partial x^i/\partial t)$ and $\theta=a(v+B)$.
Once we allow for deviations from homogeneity in the matter and metric, we should also allow for inhomogeneity in an interacting vacuum, $V(t)+\delta V(t,x^i)$. As remarked earlier, the vacuum has an energy density and pressure, but no unique velocity. In particular the momentum of the vacuum vanishes in any frame, $(\rhovac+\Pvac)\theta=0$.
On the other hand the energy flow $\uvac$ defined in Eq.~(\ref{def-uvac}), can be written in analogy with the fluid velocity (\ref{u}) as
\be
 \label{def-thetavac}
 \uvac^\mu = \left[ 1-\phi \,, a^{-1}\partial^i \vvac \right] \,, \quad \uvac_\mu = \left[ -1-\phi \,, \partial_i \thetavac \right] \,.
 \ee
where from Eq.~(\ref{def-uvac}) we identify $\thetavac=-\delta V/\dot{V}$.

Perturbations about a spatially flat ($K=0$) FRW metric \cite{Kodama:1985bj,Mukhanov:1990me,Malik:2004tf,Malik:2008im} are described by the line element
\be
 ds^2 = -(1+2\phi)dt^2+2a\partial _i B dt dx^i
  + a^2 \left[(1-2\psi)\delta_{ij}+2\partial_i\partial_j E \right] dx^i dx^j \,.
 \ee

Following \cite{Kodama:1985bj,Malik:2004tf,Malik:2008im}, we decompose the energy-flow along and orthogonal to the fluid velocity,
\be
 \label{def-f}
 Q_\mu  = Q u_\mu + f_\mu
 \,,
 \ee
where $f_\mu u^\mu=0$, so that we have
\be
 Q_\mu = \left[ -Q(1+\phi)-\delta Q \,, \partial_i (f + Q\theta) \right] \,.
 \ee
The energy continuity equations for matter and vacuum become
\bea
 \dot{\delta\rho} + 3H(\delta\rho+\delta P)
 -3 (\rho+P) \dot\psi + (\rho+P)\frac{\nabla^2}{a^2} \left( \theta + a^2\dot{E} - aB \right)
 &=& -\delta Q - Q\phi
 \,,\\
 \dot{\delta V} &=& \delta Q + Q\phi \,.
\eea
while the momentum conservation becomes
\bea
(\rho+P)\dot\theta - 3c_s^2H(\rho+P)\theta + (\rho+P)\phi
+\delta P \nonumber &=& - f + c_s^2 Q \theta
 \,,\\
 -\delta V &=& f + Q\theta \,.
\eea
where the adiabatic sound speed $c_s^2\equiv \dot{P}/\dot\rho$.
Note that the vacuum momentum conservation equation becomes a constraint equation which requires that the vacuum pressure gradient is balanced by the force
\be
 \nabla_i (-V) = \nabla_i (f+Q\theta) \,.
\ee
This determines the equal and opposite force exerted by the vacuum on the matter:
\be
- f = \delta V + \dot{V}\theta \,.
 \ee
i.e., the fluid element feels the gradient of the vacuum potential energy.

Note that the perturbations of a fluid coupled to the vacuum with $\Pvac=-\rhovac$ has no additional degrees of freedom, in contrast to a dark energy fluid with $P_X\neq -\rho_X$. Using the vacuum energy and momentum conservation equations to eliminate $\delta Q$ and $f$ we obtain
\bea
 \label{finaldeltarho}
 \dot{\delta\rho} + 3H(\delta\rho+\delta P)
 -3 (\rho+P)\dot\psi + (\rho+P) \frac{\nabla^2}{a^2} \left( \theta + a^2\dot{E} - aB \right)
 = - \dot{\delta V}
 \,,\\
 \label{finaltheta}
(\rho+P)\dot\theta - 3c_s^2H(\rho+P)\theta + (\rho+P)\phi
+\delta P = \delta V + (1+c_s^2) \dot{V} \theta
 \,.
\eea

\subsection{Gauge invariant perturbations}

It is well known that metric and matter perturbations can be gauge-dependent under a first-order gauge transformation, such as $t\to t+\delta t(t,x^i)$.
The fluid density and pressure transform as $\delta\rho\to\delta\rho-\dot\rho\delta t$ and $\delta P\to \delta P-\dot{P}\delta t$ \cite{Malik:2008im}.
Similarly the vacuum perturbation transforms as $\delta V\to \delta V -Q\delta t$ and $\delta Q\to \delta Q-\dot{Q}\delta t$. The fluid 3-momentum transforms as $\theta\to \theta+\delta t$ and the energy flow transforms similarly as $\thetavac\to\thetavac+\delta t$.

We can construct gauge-invariant perturbations by specifying quantities on a particular physical reference frame \cite{Malik:2008im}.
For example, the vacuum perturbation on hypersurfaces orthogonal to the energy transfer, $Q_\mu$, can be shown to vanish identically:
\be
 \Delta V_{\rm com} = \delta V + \dot{V} \thetavac = 0 \,,
 \ee
since from Eq.~(\ref{def-uvac}) and~(\ref{def-thetavac}) we have $\thetavac=-\delta V/\dot{V}$. This simply reflects that the energy flow is the gradient of the vacuum energy and therefore the orthogonal hypersurfaces are uniform vacuum energy hypersurfaces by construction.

On the other hand the vacuum perturbation on hypersurfaces orthogonal to the matter 4-velocity, $u^\mu$, (the {\em comoving vacuum perturbation}) is given by
\be
 \label{defdeltarhovac}
 \delta V_{\rm com} = \delta V + \dot{V} \theta = -f \,.
 \ee
This is in general non-zero, i.e., the vacuum may be spatially inhomogeneous in the comoving-orthogonal gauge.
For example, the Poisson equation for the Newtonian metric potential is given by
\be
 \label{Poisson}
\nabla^2 \Phi = 4\pi G \left( \delta\rho_{\rm com} + \delta V_{\rm com} \right) \,,
\ee
i.e., the Newtonian metric potential is sourced by both the matter and vacuum perturbations, where
\be
\delta\rho_{\rm com} = \delta\rho + \dot\rho\theta \,.
\ee

\begin{figure}
\centering
\includegraphics[width=1.2\textwidth]{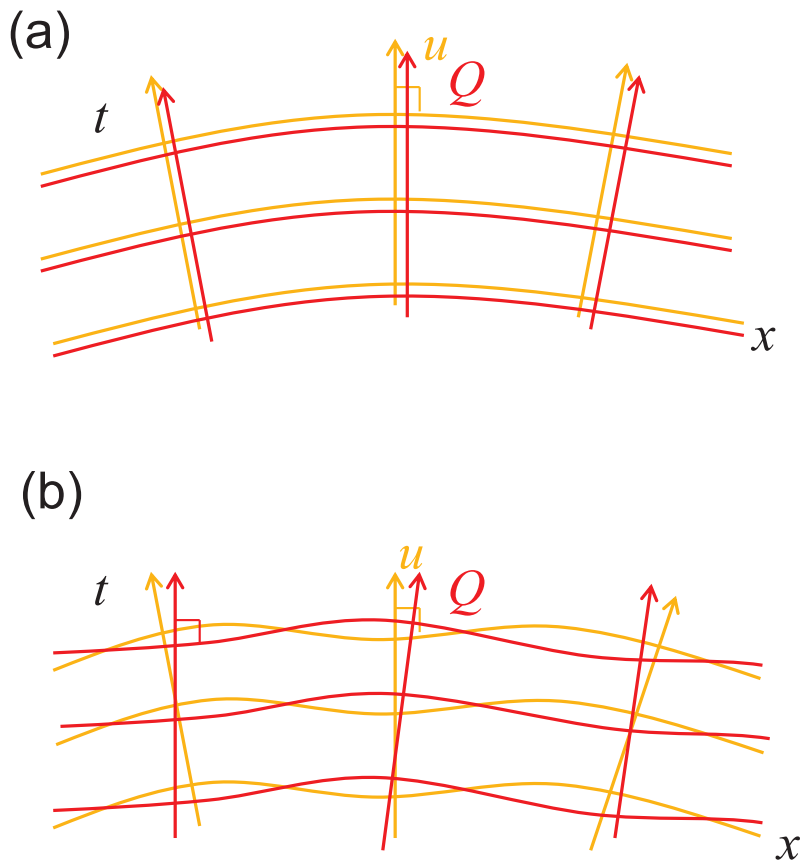}
\vspace*{-2cm}
\caption{(a) In an FRW cosmology the homogeneous spatial hypersurfaces are orthogonal to both the fluid 4-velocity, $u^\mu$, and the vacuum energy flow, $Q^\mu$. (b) In an inhomogeneous cosmology the spatial hypersurfaces orthogonal to the fluid 4-velocity (light orange) and the vacuum energy flow (dark red) do not necessarily coincide.
% The arbitrary choice of reference frame is reflected in the gauge-dependence of quantities such as the vacuum perturbation, $\delta V$.
}
\label{fig}
\end{figure}

Note that we can write the comoving vacuum density perturbation (\ref{defdeltarhovac}) as
\be
 \label{defdeltarhovac2}
 \delta V_{\rm com} = \dot{V} \left( \theta - \thetavac \right) \,.
 \ee
Therefore, if the energy flow follows the fluid four-velocity, $\uvac^\mu=u^\mu$, then we have $\thetavac=\theta$ and the vacuum is spatially homogeneous on comoving-orthogonal hypersurfaces, $\delta V_{\rm com}=0$.

Another gauge invariant expression for the vacuum density perturbation is the dimensionless vacuum perturbation on uniform-fluid density hypersurfaces, which describes a relative density perturbation
\be
 \label{defSvac}
 \Svac
% = 3 \left(\zetavac - \zeta \right)
 = -3H \left( \frac{\delta V}{\dot{V}} - \frac{\delta\rho}{\dot\rho} \right) \,.
 \ee
If, for example, the vacuum energy is a function of the local matter density, $V=V(\rho)$, then the relative density perturbation must vanish and the vacuum is spatially
% v3
homogeneous
on uniform-density hypersurfaces, $\Svac=0$.

The total non-adiabatic pressure perturbation due to any intrinsic non-adiabatic pressure of the matter and the relative entropy perturbation between the vacuum and matter is then
%, from (\ref{Pnad}) and (\ref{defSvac}),
\bea
 \label{defPnad}
 \delta P_{\rm nad}
 % = \delta P - c_s^2 \delta\rho + \frac{(1+c_s^2)\dot\rho\dot{V}}{3H(\dot\rho+\dot{V})} \Svac
%
% v3
  &=& \delta P - \delta V - \left( \frac{\dot{P}-\dot{V}}{\dot{\rho}+\dot{V}} \right) \left( \delta\rho + \delta V \right) \,, \nonumber\\
  &=& \delta P - c_s^2 \delta\rho + \frac{(1+c_s^2)Q[Q+3H(\rho+P)]}{9H^2(\rho+P)} \Svac \,,
 \eea
where the adiabatic sound speed for matter is $c_s^2=\dot{P}/\dot\rho$.
This vanishes for adiabatic matter perturbations, $\delta P=c_s^2\delta\rho$, and adiabatic vacuum fluctuations, $\Svac=0$, or a non-interacting vacuum, $Q=0$.

\section{Decomposed generalised Chaplygin gas}

As an example of how an inhomogeneous vacuum energy might be used to describe the present accelerated expansion of our Universe we will show how one widely-studied dark energy model, the Chaplygin gas, can be re-interpreted as an interacting vacuum+matter cosmology, and how this re-interpretation can motivate different possible behaviour for density perturbations.

Any
% spatially homogeneous FRW cosmology with given
dark energy fluid energy-momentum tensor (\ref{Tmunu}) with density $\rho_\de$ can be described by pressureless matter, with density $\rho_m$ and velocity $u^\mu_m=u^\mu$, interacting with the vacuum, $V$, such that $\rho_\de=\rho_m+V$ \cite{Wands:2012vg}. The corresponding matter and vacuum densities are given by
\be
 \label{decompose}
 \rho_m = \rho_\de+P_\de \,, \quad V = -P_\de \,.
 \ee
while the energy flow is $Q_\mu=\nabla_\mu P_\de$.
In an FRW cosmology this corresponds to $Q=-\dot{P}_\de$.
One might choose to decompose a dark energy model $\rho_\de(a)$ into any two interacting barotropic fluids such that $\rho_\de=\rho_1+\rho_2$, but this would double the degrees of freedom in the model unless one of these two ``fluids'' is the vacuum.

% Linder and Scherrer \cite{Linder:2008ya} showed that any dark energy cosmology could be represented by a constant vacuum energy, $V$, and a fluid with an suitable equation of state, $w_f=(P_{de}+V)/(\rho_{de}-V)$. In this case the vacuum energy was non-interacting but the fluid required a specific equation of state, whereas in our case we will consider pressureless matter, but the vacuum then requires a specific interaction.\footnote{In practice one could choose any equation of state for the matter and a specific interaction such that $\rho_m+V=\rho_{de}$. An example is a quintessence model, where a canonical scalar field has sound speed equal to one, which reproduces any given dark energy cosmology for a suitably chosen potential.} These are both examples of so-called dark degeneracy~\cite{Kunz:2007rk,Aviles:2011ak}.

The generalised Chaplygin gas, defined by the barotropic equation of state \cite{Kamenshchik:2001cp,Bento:2002ps}
\be
 \label{Pgcg}
 P_\gcg = -A \rho_\gcg^{-\alpha} \,.
 \ee
This leads to a solution for the density in an FRW cosmology
\be
 \rho_\gcg = \left( A + Ba^{-3(1+\alpha)} \right)^{1/(1+\alpha)} \,.
 \ee
This has the simple limiting behaviour $\rho_\gcg\propto a^{-3}$ as $a\to0$ and $\rho_\gcg\to A^{1/(1+\alpha)}$ as $a\to+\infty$, therefore this has been proposed as a unified dark matter model. However such models are strongly constrained by observations since the barotropic equation of state defines a sound speed for matter perturbations which only reproduces the successful $\Lambda$CDM model when $\alpha\to0$ \cite{Sandvik:2002jz,Park:2009np}.

The decomposition (\ref{decompose}) into pressureless matter interacting with the vacuum has previously been considered for the generalised Chaplygin gas by Bento et al \cite{Bento:2004uh}. In this case
we have the FRW solution
 \be
 V =
%  A \rho_\gcg^{-\alpha}
 A \left( A + Ba^{-3(1+\alpha)} \right)^{-\alpha/(1+\alpha)}
%  \,, \quad \rho_m = \rho_\gcg +P_\gcg
   \,,
\ee
and hence
\be
\label{defA}
 A = (\rho_m + V)^\alpha V \,.
 \ee
% and the energy transfer is given by
% \be
% Q = -\dot{P}_\gcg = -\alpha H A \rho_\gcg^{-\alpha}
% \ee
The form of the FRW solution suggests a simple interaction
\be
 \label{Qgcg}
 Q
= 3 \alpha H \left( \frac{\rho_m V}{\rho_m + V} \right) \,.
 \ee
In the matter or vacuum dominated limits this reduces to an interaction of the form $Q\propto H\rho_m$ or $Q\propto HV$ studied, for example, by Barrow and Clifton~\cite{Barrow:2006hia}.

It is intriguing to note that the FRW solution can be defined in terms of an interaction (\ref{Qgcg}) with a single dimensionless parameter, $\alpha$, whereas when defined in terms of an equation of state (\ref{Pgcg}) its definition requires both $\alpha$ and the dimensional constant $A$ which determines the late-time cosmological constant. In the interaction model, $A$ (and therefore the late-time cosmological constant) emerges as an integration constant dependent on initial conditions.

However, to study inhomogeneous perturbations in the decomposed model we must ``lift'' the explicitly time-dependent FRW solution to a covariant model for the interaction. We have at least two choices.
In either case one can calculate the speed of sound in the combined interacting vacuum+matter using the usual definition \cite{ArmendarizPicon:1999rj,Weller:2003hw,Bean:2003fb,Hannestad:2005ak} and the decomposed density and pressure (\ref{decompose})
\be
 \label{cint}
 c_{\rm de}^2 \equiv \left( \frac{\delta P_{\rm de}}{\delta\rho_{\rm de}} \right)_{\rm com}
  = \left( \frac{- \delta{V}}{\delta\rho_m+\delta{V}} \right)_{\rm com}\,.
 \ee

\subsection{Barotropic model}

Firstly one could require that the local vacuum energy is a function of the local matter density, $V=V(\rho_m)$. If this applies to inhomogeneous perturbations of the interacting vacuum+matter as well as the background then we require
\be
 \label{adiabatic}
 \delta V = \frac{\dot{V}}{\dot{\rho}_m}\delta\rho_m \,.
 \ee

This implies that the vacuum perturbations are adiabatic and $\Svac=0$ in Eq.~(\ref{defSvac}). On the other hand the comoving vacuum energy, (\ref{defdeltarhovac}), is non-zero
\be
 \delta V_{\rm com}
 % = \delta V + \dot{V}\theta
 = \frac{\dot{V}}{\dot{\rho}_m} \delta\rho_{m,{\rm com}} \,.
 \ee
Thus one needs to consistently include the vacuum inhomogeneities as well as matter inhomogeneities, for example in the Poission equation (\ref{Poisson}),  when $\dot{V}\neq0$.

Using the adiabatic condition (\ref{adiabatic}) we obtain the dark energy sound speed (\ref{cint})
\be
 c_{\rm int}^2 = \frac{\dot{P}_{\rm gCg}}{\dot{\rho}_{\rm gCg}} \,.
 \ee
We see therefore the the sound speed for the interacting vacuum+fluid is the same as the adiabatic sound speed of the original barotropic Chaplygin gas (\ref{Pgcg}). Perturbations of the the interacting vacuum+matter respect the same barotropic equation of state as the original fluid (\ref{Pgcg}) and therefore the observational predictions for the cosmic microwave background (CMB) anisotropies, for example, are exactly the same as the original fluid model \cite{inprep}.

CMB temperature anisotropies as well as the power spectrum for the interacting vacuum+matter are shown in figures~2 and~3. We see that only very small values of the dimensionless parameter $|\alpha|<10^{-4}$ are allowed without introducing unacceptably large oscillations in the matter power spectrum \cite{Sandvik:2002jz,Park:2009np}, forcing the model to be extremely close to the standard $\Lambda$CDM cosmology.

\begin{figure}
\centering
 \includegraphics[width=0.8\textwidth]{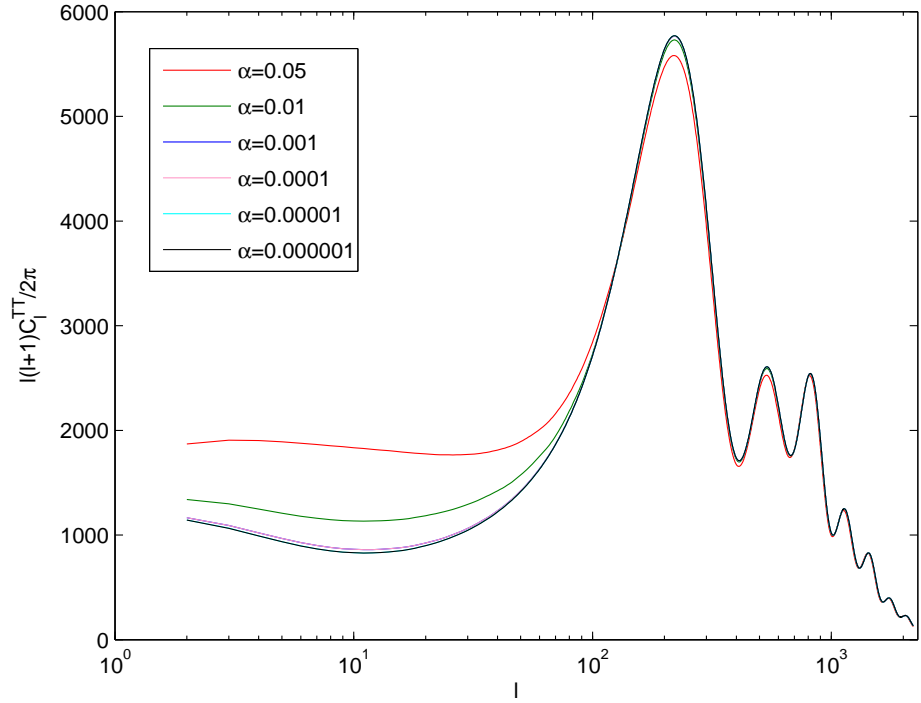}
% \vspace*{-2cm}
\caption{Cosmic microwave background angular power spectrum for the generalised Chaplygin gas with barotropic equation of state and an adiabatic sound speed~\cite{inprep}.
}
\label{fig-gcgcl}
\end{figure}

\begin{figure}
\centering
 \includegraphics[width=0.8\textwidth]{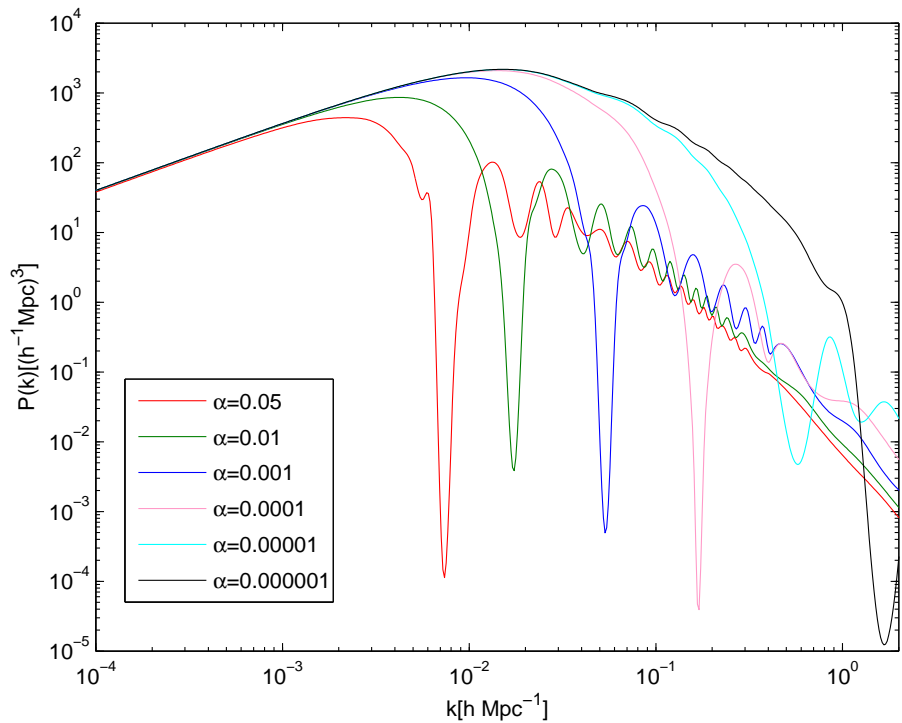}
% \vspace*{-2cm}
\caption{Power spectrum for baryons plus generalised Chaplygin gas with barotropic equation of state and an adiabatic sound speed~\cite{inprep}.
}
\label{fig-gcgpk}
\end{figure}

\subsection{Geodesic flow}

The interacting vacuum+matter model allows a different behaviour for inhomogeneous perturbations than the barotropic fluid. Suppose that the energy flow, $Q_\mu$, is along the matter 4-velocity, $u_\mu$. The force exert exerted by the vacuum on the matter, $-f_\mu$ in Eq.~(\ref{def-f}), is zero and matter 4-velocity is a geodesic flow.

In this case the comoving vacuum perturbation, $\delta V_{\rm com}$ in Eq.~(\ref{defdeltarhovac}), is zero, i.e., the vacuum is spatially homogeneous on hypersurfaces orthogonal to the matter 4-velocity. This means that the comoving pressure perturbation for the interacting vacuum+matter is zero and hence the sound speed ({\ref{cint}) is zero, $c_{\rm int}^2=0$~\cite{Creminelli:2008wc,Lim:2010yk}.

The evolution of density perturbations is therefore much closer to that in a standard $\Lambda$CDM cosmology with zero sound speed than the original Chaplygin gas model with $\alpha\neq0$. CMB temperature anisotropies as well as the matter power spectrum are shown in figures~4 and~5. Much larger values of $\alpha\sim0.1$ may be compatible with the data in this case \cite{inprep}.

Note however that this geodesic model for the interaction does allow a gauge-invariant vacuum entropy perturbation~(\ref{defSvac})
\be
 \Svac = \frac{3H}{\dot{\rho}_m} \delta\rho_{m,{\rm com}} \,.
 %= - \frac{(\rho_m+V)^2}{\alpha\rho_m\left(\rho_m+(1+\alpha)V\right)}\frac{\delta A}{A} \,,
 \ee
%Hence from Eq.~(\ref{Pgcg}) we can identify the relative entropy perturbation with a change in the local equation of state and non-adiabatic pressure perturbation (\ref{defPnad}), $\delta P_{\rm nad}=-V\delta A/A$.
%
However due to the Poisson constraint equation (\ref{Poisson}) the comoving density perturbation vanishes on large scales and hence this is compatible with adiabatic initial conditions for the interacting vacuum+matter at early times.

The requirement that $Q_\mu=Qu_\mu$ does impose an important physical restriction on the matter 4-velocity since the energy flow must then be irrotational, i.e., $u_\mu\propto \nabla_\mu V$. This would have important consquences for non-linear structure formation, possibly disastrous if gravitationally collapsed dark matter halos could not be supported by rotation. Hence we only expect geodesic matter interacting with the vacuum to be a viable model for the initial stages of structure formation and we would need a microphysical model for gravitationally collapsed halos.

%And we do not have a clear covariant model for the interaction $Q=-u^\mu Q_\mu$. By requiring $\delta V = -Q\theta$, we construct the vacuum perturbation, $\delta V$, iteratively given the matter velocity, $\theta$, and the background energy transfer, $Q$ in Eq.~(\ref{Qgcg}).

\begin{figure}
\centering
 \includegraphics[width=0.8\textwidth]{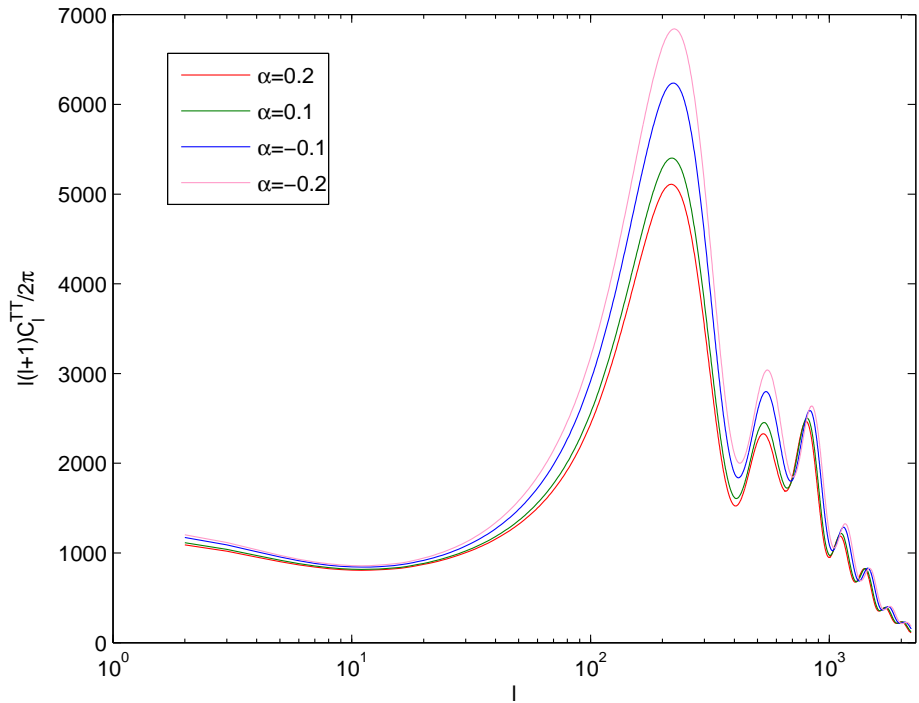}
% \vspace*{-2cm}
\caption{Cosmic microwave background angular power spectrum for the decomposed generalised Chaplygin gas with geodesic matter and zero sound speed~\cite{inprep}.
}
\label{fig-geocl}
\end{figure}

\begin{figure}
\centering
 \includegraphics[width=0.8\textwidth]{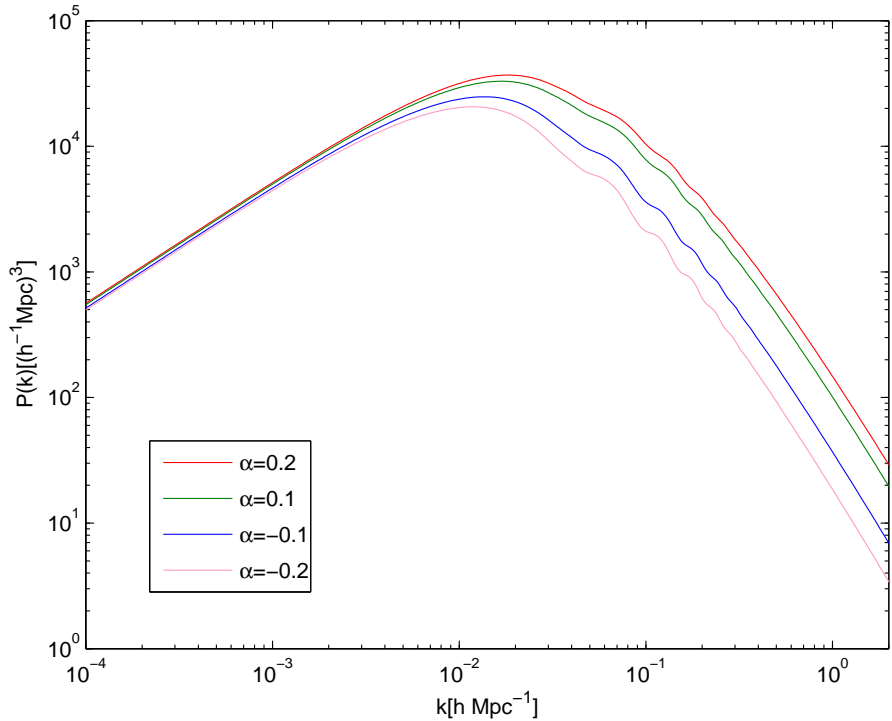}
% \vspace*{-2cm}
\caption{Power spectrum for baryons plus the decomposed generalised Chaplygin gas with geodesic matter and zero sound speed~\cite{inprep}.
}
\label{fig-geopk}
\end{figure}

\section{Discussion}
\label{sec:conclusions}

Many different explanations have been proposed for the present-day
acceleration of the Universe \cite{Copeland:2006wr}. Perhaps the most common is a
self-interacting scalar field, $\varphi$, whose self-interaction
potential energy could dominate the present energy density.
Different models are defined by the chosen functional form for the
potential energy, $V(\varphi)$ \cite{Ratra:1987rm,Caldwell:1997ii},
or the kinetic energy as a function of the field gradient,
$X=(\nabla\varphi)^2$
\cite{ArmendarizPicon:1999rj,ArmendarizPicon:2000ah}. Alternative
explanations include fluid models, specified by a barotropic
equation of state, $P(\rho)$
\cite{Kamenshchik:2001cp,Linder:2008ya,Balbi:2007mz}, some of which
could provide unified dark matter models capable of explaining
acceleration on large scales and galactic dynamics (as dark matter)
on much smaller scales. Acceleration requires an exotic equation of
state with negative pressure, and yet the sound speed must remain
real in order to avoid instabilities with respect to inhomogeneous
perturbations. More sophisticated models have been proposed which
allow for interactions between the different fields or fluid
components, e.g., coupled quintessence
models~\cite{Wetterich:1994bg,Amendola:1999er,Holden:1999hm,Billyard:2000bh,Farrar:2003uw,Boehmer:2008av}.
Such models are motivated by astronomical observations
\cite{Copeland:2006wr}, providing increasingly detailed constraints
on the model parameters, but the models themselves lack a persuasive
underpinning physical motivation.
One might therefore simply consider the simplest possible model, or parameterisation, compatible with the data.

In this paper we have considered vacuum energy as a source of spacetime curvature in Einstein gravity. A homogeneous vacuum in Einstein gravity is equivalent to a cosmological constant, whereas an inhomogeneous (time- or space-dependent) vacuum implies an interacting vacuum.
%
%The vacuum energy is a scalar potential and thus defines a particular spacetime foliation. In a cosmological solution this implies that there exist spatial hypersurfaces on which the vacuum energy is homogeneous, but these need not coincide with, for example, comoving-orthogonal hypersurfaces when one considers inhomogeneous perturbations and structure formation in our Universe.
%
Simply specifying a particular time-dependence for the vacuum energy in an FRW cosmology is unsatisfactory if the vacuum energy is introduced solely to produce a particular time-dependence of the cosmological expansion. Different physical models for the origin of the vacuum energy, or its interaction with other matter fields, will lead to different cosmological behaviour. Even models which yield the same FRW background solutions \cite{Kunz:2007rk,Aviles:2011ak} may be distinguished for instance by the predictions for CMB anisotropies or the matter or galaxy power spectrum.

As an example, we have shown how the generalised Chaplygin gas cosmology can be re-derived as a solution to an interacting matter+vacuum model. The interaction can be defined by a single dimensionless constant and the late-time constant vacuum energy (which appears as a dimensional parameter in the original Chaplygin gas model) can instead be derived as a constant of integration in the matter+vacuum model. We have shown the CMB and matter power spectrum predictions for two models for interacting vacuum+matter which yield the same background solution as the generalised Chaplygin gas, but give very different observational predictions \cite{inprep}.

Another familiar example, which we have not discussed here, would be a quintessence model with a self-interacting scalar field, $\varphi$. The self-interaction potential of the field, $V(\varphi)$, provides a vacuum energy density interacting with the kinetic energy of the scalar field. It is well known that a canonical scalar field has a sound speed equal to unity and one might thereby hope to distinguish it from a barotropic fluid model where the sound speed is necessarily equal to the adiabatic sound speed. By considering a broader class of interacting vacuum models one can study models with a range of different possible sound speeds. Our equations enable us to consider vacuum energy interacting with other forms of matter, including pressureless matter or radiation. We can therefore consider vacuum energy models which do not introduce any degrees of freedom beyond those already present in the cosmology, e.g., unified dark energy models.

In our work we have considered only linear perturbations but non-linear evolution may provide further observational constraints on different models. Often it is assumed that the vacuum energy remains unperturbed, or negligible, during collapse, but some cases, such as the interacting vacuum+matter with geodesic 4-velocity and zero sound speed may be amenable to a study of non-linear collapse \cite{Creminelli:2009mu} and we hope to study this further in future.

\bigskip
% \acknowledgments
{DW is grateful to the organisers of the IVth International Conference on Gravitation and Cosmology, Guadalajara, for their hospitality.
%We thank Marco Bruni, Rob Crittenden, Georgia Kittou, Kazuya Koyama and Roy Maartens for helpful comments.
DW is supported by STFC grant ST/H002774/1. JD-S is supported by CONACYT grant 210405. JD-S and YW thank the ICG, University of Portsmouth for their hospitality.}

\end{document}